\documentclass[sigconf,screen]{acmart}

\AtBeginDocument{%
  \providecommand\BibTeX{{%
    \normalfont B\kern-0.5em{\scshape i\kern-0.25em b}\kern-0.8em\TeX}}}

\usepackage{ragged2e}
\begin{document}

\author{Francesco Pierri}
\affiliation{%
  \institution{Politecnico di Milano, Milano, Italy}
    \country{}
  }

\title{Political advertisement on Facebook and Instagram in the run up to 2022 Italian general election}

\begin{abstract}
Targeted advertising on online social platforms has become increasingly relevant in the political marketing toolkit. Monitoring political advertising is crucial to ensure accountability and transparency of democratic processes. Leveraging Meta public library of sponsored content, we study the extent to which political ads were delivered on Facebook and Instagram in the run-up to the 2022 Italian general election. Analyzing over 23 k unique ads paid by 2.7 k unique sponsors, with an associated amount spent of 4 M EUR and over 1 billion views generated, we investigate temporal, geographical, and demographic patterns of the political campaigning activity of main coalitions. We find results that are in accordance with their political agenda and the electoral outcome, highlighting how the most active coalitions also obtained most of the votes and showing regional differences that are coherent with the (targeted) political base of each group. Our work raises attention to the need for further studies of digital advertising and its implications for individuals' opinions and choices.
\end{abstract}



\maketitle

\section{Introduction}
Online social media provide a quick and effective way to deliver advertisements to a target audience specified by the sponsoring entity \cite{dommett2019political}. Unlike television, where ads are broadcasted to the viewers, sponsored content on online social platforms can be delivered exclusively to the preferred audience \cite{boerman2017online}.

This feature becomes particularly relevant during political campaigns when parties and candidates are willing to reach potential voters more precisely \cite{fowler2021political}. Given the possibility to target a fine-grained audience and the low costs of setting up ads, social media advertising has become an essential part of the political marketing toolkit \cite{kreiss2016prototype}.

Nevertheless, while televised advertising has been largely studied \cite{ridout2021influence}, gaps remain in our knowledge of whether digital advertising can affect voting intentions \cite{coppock2022does,aggarwal20232}. As the amount spent on digital advertising becomes more relevant than televised ads, there are still many questions left to answer on the actual effects of political advertising delivered on online social networks, including the extent to which ads can be used to mislead online users \cite{caetano2022characterizing,martins2022characterizing}.

Social media platforms have increasingly been under scrutiny for providing a channel for divisive and controversial messaging \cite{entous2017russian}. As a response to these criticisms, and following the events of Cambridge Analytica in 2018\footnote{\url{https://en.wikipedia.org/wiki/Facebook–Cambridge_Analytica_data_scandal}}, Meta\footnote{Formerly known as Facebook.} launched its Ad Library platform, giving public access to sponsored content on its platforms and providing researchers with unprecedented opportunities to study political communications \cite{leerssen2021news}. Existing studies leveraged this library to study digital advertising in political settings, from analyzing the issue of migrations in Italy \cite{capozzi2020facebook} to marketing campaigns during the COVID-19 pandemic \cite{mejova2020covid,silva2021covid}. In the present work, we analyze digital advertising on two mainstream social media platforms, namely Instagram and Facebook that count around 28-35 M users in Italy\footnote{\url{https://www.statista.com/statistics/787390/main-social-networks-users-italy/}}, in the run-up to the 2022 Italian general election that took place on September 25th, 2022.

The fall of the Italian government of national unity on July 21st, 2022 led to the first-ever snap election taking place in the fall\footnote{\url{https://en.wikipedia.org/wiki/2022_Italian_government_crisis}}, with a reduced number of seats in the House of Chambers (from 630 to 400) and the Senate (from 315 to 200) due to the 2020 Italian constitutional referendum\footnote{\url{https://en.wikipedia.org/wiki/2020_Italian_constitutional_referendum}}.
The election saw a record-low voter turnout (less than 64\% of the eligible voters) and it was won by far by the right-wing coalition guided by Giorgia Meloni (over 43\% of the vote share). The Centre-left coalition obtained approximately 25\% of the voters, whereas the populist Movimento 5 Stelle reached less than 16\%. The fourth largest group was the liberal and centrist Third Pole, which included Matteo Renzi's party Italia Viva, that obtained almost 8\% of the vote share\footnote{\url{https://en.wikipedia.org/wiki/2022_Italian_general_election}}.

In this setting, we aim to address the following research questions:\\
\textbf{RQ1}: What were the reach and amount spent on political advertisement on Meta platforms in the run-up to the 2022 Italian general election?\\
\textbf{RQ2}: What were the main differences in the advertising campaigns of political groups?

We access Meta Ad Library through its API and retrieve all ads that were related to the election using a list of relevant keywords, resulting in a collection of over 23 k unique ads posted by 2.7 k unique sponsors, for a total amount of 4 M EUR spent to generate over a billion impressions. We first analyze the distributions of money spent, impressions generated, and number of ads created and active throughout the period of analysis, highlighting a significantly increasing trend of expenses and reach towards election day. We find a strong correlation between ads expenses and the number of views they generate, with an estimated average cost per thousand impressions of 4 EUR. We further focus on the advertising campaigns of the main political parties by manually labeling sponsors that are affiliated with political parties, candidates and elected representatives of the four main coalitions mentioned above. Leveraging detailed information present in our data from a temporal, geographical, and demographic perspective, we show similarities and differences which reflect the electoral weight of each group, as well as their agenda and targeted voters. Until recently, there were limited resources to study advertising campaigns on social media platforms. Showing the potentialities offered by a large-scale data collection tool such as Meta Ad library, our study contributes to the existing literature on the role of social media in influencing individuals' opinions and actions in the real world, calling for further research on the effects of targeted advertising. 

The outline of this paper is the following: in the next section we review the existing literature related to our work; then, we describe the methodologies employed to collect and analyze the data; next, we provide results to address our research questions. Finally, we discuss our findings, mention limitations to our analysis and draw future work.

\begin{figure}[!t]
    \centering
   \includegraphics[width=0.85\linewidth]{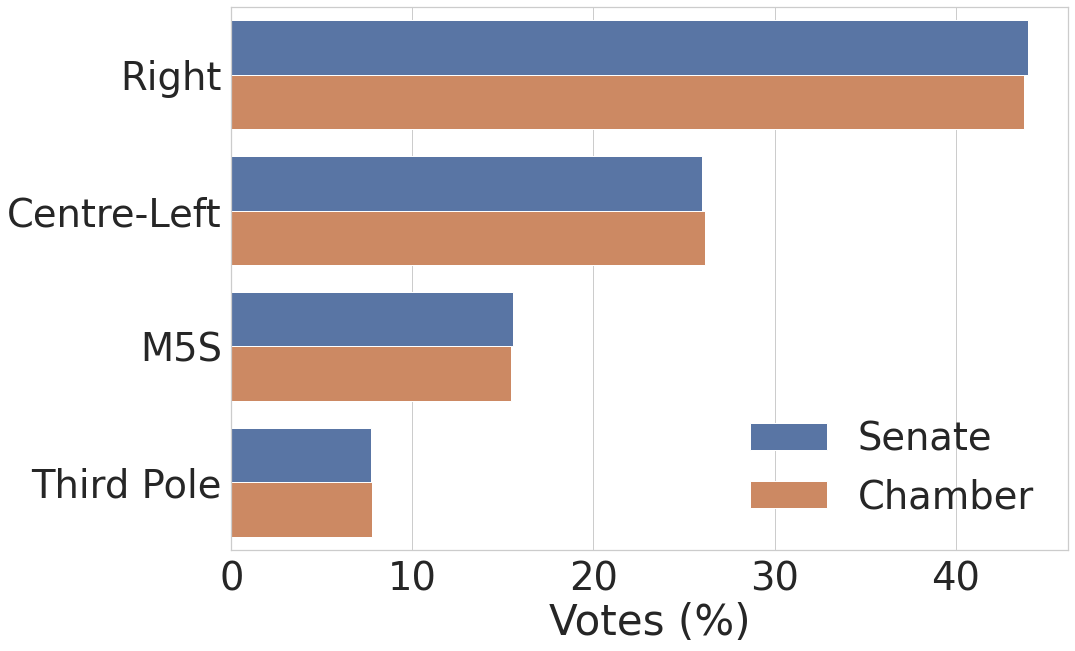}
    \caption{Electoral results obtained by each political coalition for the Senate and the Chamber of deputies.}
    \label{fig:votes}
\end{figure}

\section{Related Work}
\citet{ribeiro2019microtargeting} study malicious advertising put in place by the Russian Intelligence Research Agency (IRA) prior to the 2016 U.S. elections. Leveraging a dataset of 3,517 Facebook advertisements released by the Democrats Permanent Select Committee on Intelligence, and conducting surveys, they look at how users with different political ideologies report, approve, and perceive truth in the content of the IRA ads.

\citet{mejova2020covid} use Meta (formerly Facebook) Ad Library to study advertising campaigns during the COVID-19 pandemic, finding that the crisis is used to advertise political attacks, donation solicitations, business promotion, stock market advice, and animal rights campaigning. They also show the presence of misinformation about the virus, ranging from bioweapons conspiracy theories to unverifiable claims by politicians.

\citet{capozzi2020facebook} aim to study micro-targeting in the political messaging around migration in Italy, collecting data from Meta Ad Library. They find that different parties focus on different demographic cohorts, and show that nationalist parties frequently target male audience in their migration ads.  

\citet{silva2020facebook} put in place an independent auditing system to monitor political ads on Facebook in the run-up to the 2018 Brazilian elections. Gathering ads from the timeline of volunteering Facebook users, they employ a set of supervised machine learning models to automatically classify political ads, noticing that not all of them were labeled as political from Facebook.

\citet{silva2021covid} also investigate  the implications of the COVID-19 pandemic by collecting Facebook ads (using a combination of API and crawlers) sponsored in Brazil, finding evidence of a huge amount of advertisements promoting or denigrating the image of political actors, as well as health-related misinformation (such as the use of Hydroxychloroquine for COVID-19 treatment).

\begin{figure}[!t]
    \centering
   \includegraphics[width=\linewidth]{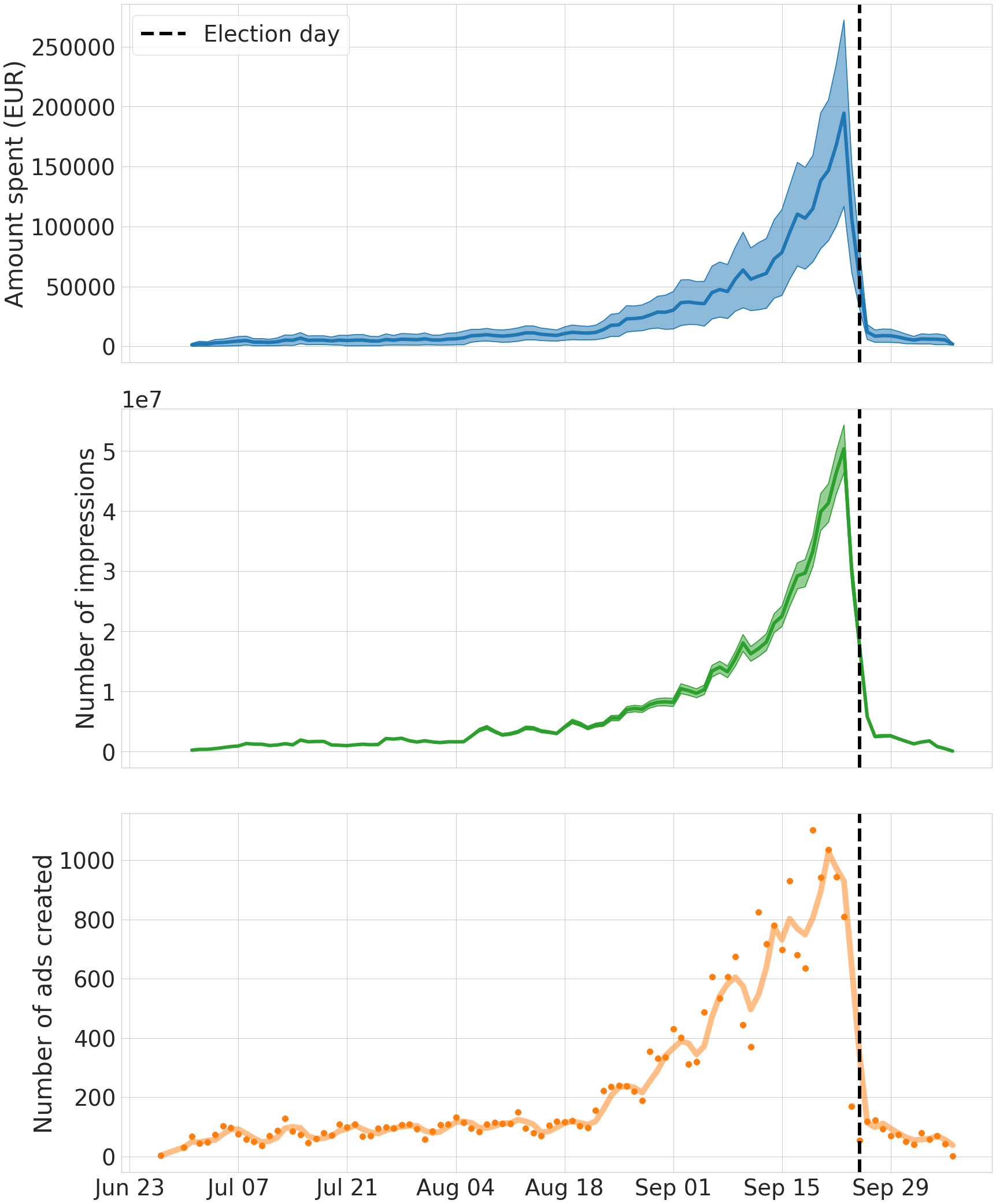}
    \caption{Time series of the total amount of EUR spent, the number of impressions generated and the number of unique ads created each day in our dataset. In the bottom panel the solid line represents a 3-day moving average. In the two upper panels the dashed area represents lower and upper bounds as provided by Meta Ad Library API, with the solid line corresponding to the mean value.}
    \label{fig:ts-general}
\end{figure}

\citet{calvo2021global} and \citet{baviera2022political} study the advertising strategies of national parties in the run-up to general elections held in Spain in 2019. To this end, they analyze a corpus of over 14 k Facebook ads collected from Meta Ad Library and provide comparative descriptive statistics of the political communication of different parties.

\citet{capozzi2021clandestino} study migration-related advertising campaigns in Italy over one year by collecting over 2 k Facebook ads from Meta Ad Library. They build a pro-/anti- immigration classifier to label these ads, revealing a strong partisan divide among the major Italian political parties, with anti- immigration ads accounting for nearly 15 M impressions. 

\citet{le2022audit} quantify whether Facebook correctly identifies political ads and ensures compliance by advertisers, by performing a large-scale analysis of 4.2 million political and 29.6 million non-political ads from 215,030 advertisers collected from Meta API. Using a crowd-sourced approach, they show that current enforcement is very imprecise, as several advertisers are able to run political ads without disclosing them and while they are temporarily prohibited. They also highlight main gaps in the enforcement system and provide a set of recommendations to improve the security of the online political ad ecosystem.

\begin{figure*}[!t]
    \centering
   \includegraphics[width=\linewidth]{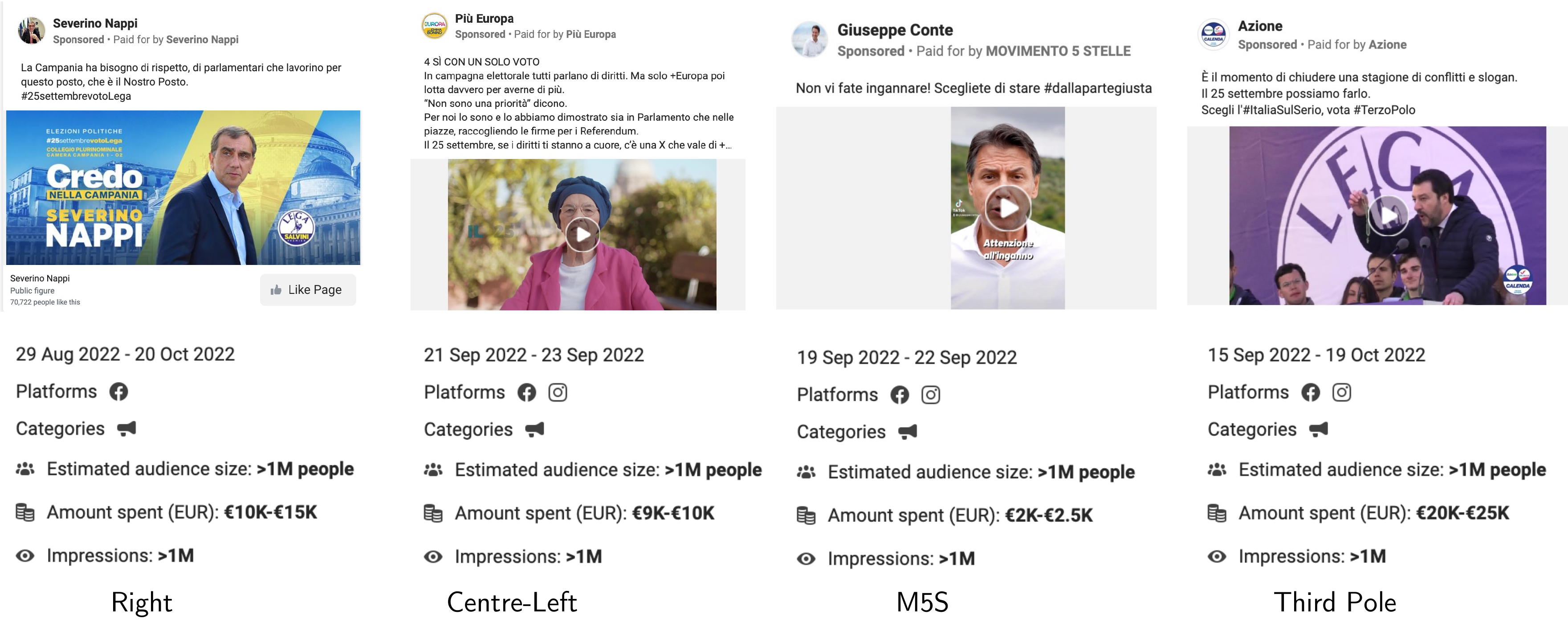}
    \caption{Most expensive ad for each political coalition as available in Meta Ad Library interactive search console.}
    \label{fig:top-ads}
\end{figure*}

\section{Data and Methods}
\subsection{Ads collection}
We collected all ads about "social issues, elections or politics" that were active on Meta platforms (Facebook, Instagram, Messenger, and the Audience Network) in the period July 1st, 2022 - October 7th, 2022, using Meta Ad Library API\footnote{\url{https://www.facebook.com/ads/library/api}}. This period was chosen because a political crisis in the former Italian government started at the beginning of July, with PM Mario Draghi eventually resigning on July 21st. The election was then held on September 25th, and our data collection captures also the pre-crisis period.

Meta requires advertisers to confirm their identity and location and declare who funded the ads\footnote{\url{https://www.facebook.com/business/help/198009284345835?id=288762101909005}}, and the API interface allows to search the entire collection of ads ran on Meta platforms.
Specifically, we queried the API with over 60 keywords in the Italian language related to the 2022 Italian election. These were obtained with a snowball sampling approach and are presented by \citet{pierri2023ita}; a sample can be found in Table \ref{tab:keywords}. The API only allows to search ads matching the query using one keyword at a time, and we eventually aggregated separate outputs discarding duplicated ads appearing in multiple searches. We provide access to the ID of ads, which can be retrieved through Meta Ad Library interactive search console and API, in the repository associated with \cite{pierri2023ita}\footnote{\url{https://github.com/frapierri/ita-election-2022}}.

\begin{table}[!t]
\centering
\begin{tabular}{llll}
elezioni & partito democratico & berlusconi & iovotolega \\ \hline
renzi & movimento 5 stelle & salvini & fratelli d'Italia \\ \hline
calenda & di maio & politiche2022 & 25settembre \\ \hline
meloni & elezioni2022 & conte & iovotoitaliaviva \\ \hline
\end{tabular}
\caption{A sample of Italian language keywords related to the 2022 election that were used to query Meta Ad Library.}
\label{tab:keywords}
\end{table}

The resulting dataset contains 23,545 unique ads posted by 2,698 unique sponsors. For each ad, the API returns several attributes, including date of creation, the period when the ad is active, name of the sponsor, message, platform on which the ad is active, intervals for the amount spent and the number of impressions generated, etc\footnote{More details are available on the Meta Ad Library webpage.}. Since Meta Ad Library provides intervals for the number of impressions generated (e.g. $[0, 999]$) and EUR spent (e.g. $[0, 99]$), for the rest of the paper we consider the mean value of such intervals when we analyze expenses and impressions. Following this rationale, we estimate that in total 4 M EUR were spent to generate over a billion impressions. We remark that the number of impressions does not correspond to the number of unique individuals that saw the ad, i.e., the same user might have seen the advertisement multiple times. We will use interchangeably "views" and "impressions" throughout the text.

Over 50\% of the ads were active on both Facebook and Instagram\footnote{Other platforms accounted for a dozen ads in total.}, whereas approximately 40\% and 6\% were active only on Facebook and Instagram, respectively. We analyzed them aggregately, and leave a comparative analysis of the two platforms for future research.

\subsection{Political labeling}
To study the advertising campaigns of political groups, we first automatically matched sponsors' names against a list of over 7,000 candidates, which is publicly available here\footnote{\url{https://github.com/ondata/elezioni-politiche-2022}}. We further manually matched sponsors with at least 10 ads present in the dataset (approximately 1,000) to candidates and parties as well as other politicians explicitly supporting the coalition during the electoral campaign\footnote{For instance Alfonso Pecoraro Scanio, who is not officially part of the M5S but that has been officially supporting the party over the last years.}, considering only that could be explicitly linked to the following political coalitions/parties:
\begin{itemize}
    \item Right (parties: Fratelli d'Italia, Lega, Forza Italia, Noi Moderati) 
    \item Centre-Left (parties: Partito Democratico, +Europa, Alleanza Verdi e Sinistra, Unione Civica)
    \item M5S (party: Movimento 5 Stelle)
    \item Third Pole (parties: Italia Viva, Azione)
\end{itemize}
We noticed some cases where ads were published by a candidate’s page but sponsored by a different entity; we matched these entities to the coalition of the sponsored candidate(s) as we never encountered an entity sponsoring candidates from different coalitions.
We refer the reader to \cite{bobbio1996left} for a definition of left-wing and right-parties in the Italian context which compose the two main coalitions; according to \cite{rooduijn2019populist} Movimento 5 Stelle is to be considered as a populist party, whereas the Third Pole is a liberal and centrist group\footnote{\url{https://en.wikipedia.org/wiki/Action_–_Italia_Viva}}. We remark that in the Italian electoral system, different parties can form coalitions and run together for election. Also, there is no difference in the election process between the two chambers. We provide a breakdown of the electoral results in Figure \ref{fig:votes}.
We excluded groups with no elected representatives, and 6 candidates elected in the autonomous regions of Trentino-Alto-Adige and Valle D'Aosta. At the end of the matching procedure, we obtained 617 unique sponsors, with 171 out of 483 elected representatives\footnote{At the time of the analysis not all 600 elected representatives were available in the list.}, that were responsible for 12,367 unique ads with an associated amount spent of approx 2.8 M EUR and over 840 M impressions generated. As an illustrative example, we show in Figure \ref{fig:top-ads} a snapshot of the most expensive sponsored content paid by each political coalition.

\section{Results}
\subsection{General statistics of ads and sponsors during the electoral campaign}

We first study the amount of money spent on political advertising and the resulting number of impressions generated over time at the ad/sponsor level, considering all the ads collected in our dataset. Throughout the results, we will report p-values significance at $\alpha=.05$.

As shown in Figure \ref{fig:ts-general}, both the aggregated daily amount spent and number of impressions generated exhibit a significant increasing trend (Mann-Kendall test, $P < .001$) towards election day (September 25th), and they are also strongly correlated (Pearson $R=0.99$, $P < .001$ ). There are no laws that regulate election silence, which imposes a ban on TV/Radio campaigning on the day before the election, on social media but we do notice that political advertising on Facebook and Instagram was strongly reduced two days before election day.

\begin{figure}[!t]
    \centering
   \includegraphics[width=\linewidth]{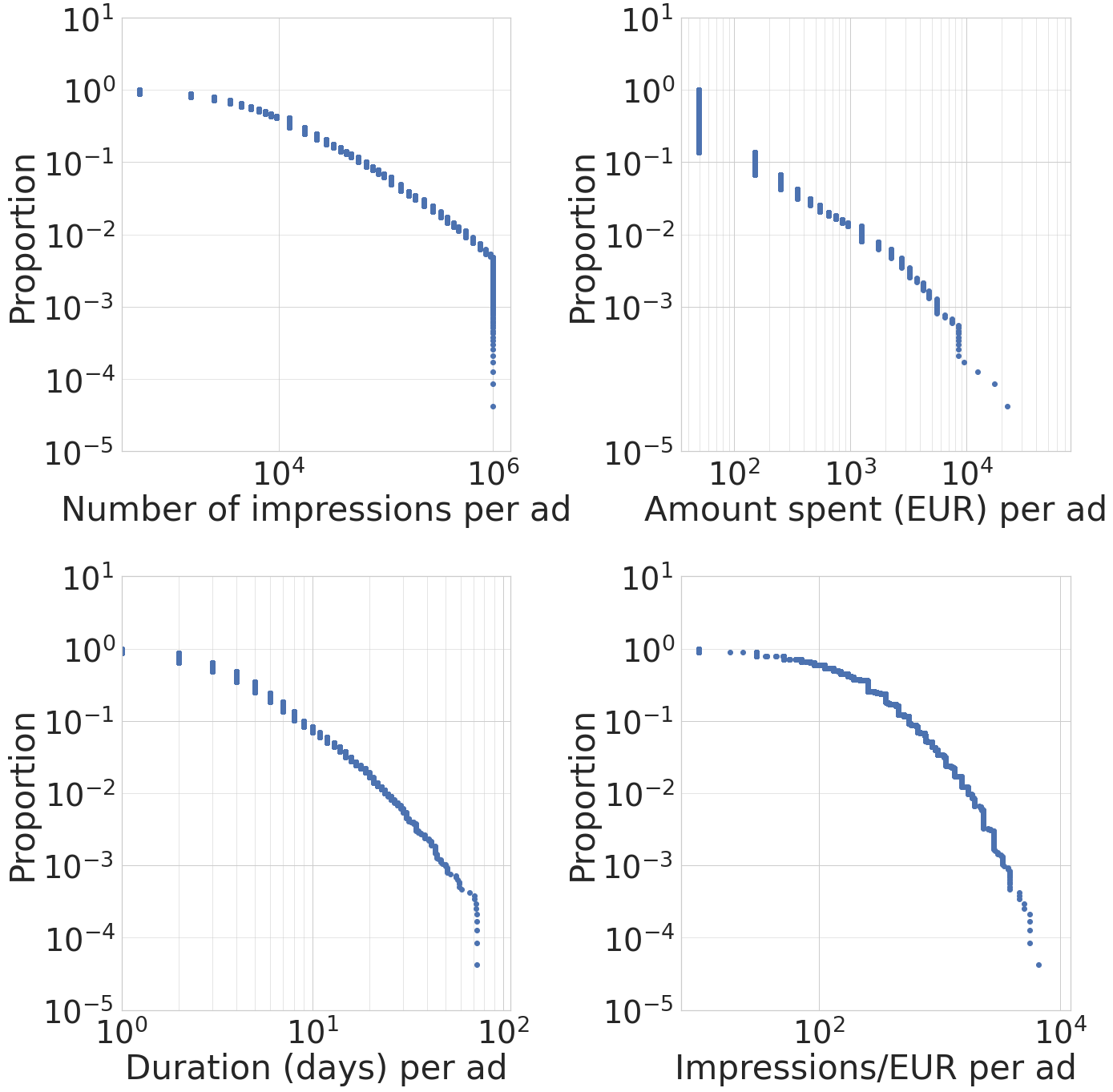}
    \caption{Distributions of the number of impressions, the amount spent, duration, and impressions/EUR for each ad in the dataset. Both axes are in logarithmic scale.}
    \label{fig:ad-distributions}
\end{figure}

Focusing on statistics of individual ads, we notice that the average amount of impressions generated is approximately 32 k (median: 7.5 k impressions), with the most performing ad generating 1 M impressions. The average expense is 111 EUR (median: 50 EUR), with the most expensive ad costing over 50 k EUR. Ads last on average 4.5 days (median: 3 days), with the longest ad lasting up to 73 days.  Finally, computing the ratio of impressions and expense, we can say that, on average, 1 EUR generated around 250 impressions (median: 131 impressions), with the most effective ad reaching over 8.5 k views for each spent EUR. All metrics exhibit an exponential-like distribution, as shown in Figure \ref{fig:ad-distributions}. Expenses are strongly correlated with the number of impressions generated (Pearson $R=0.58$, $P < .001$ ), and to a minor extent with the duration of the ad, (Pearson $R=0.19$, $P < .001$ ); duration is also positively correlated with the number of impressions generated (Pearson $R=0.3$, $P < .001$). The number of unique ads created on a daily basis exhibits a significant increasing trend (Mann-Kendall test, $P < .001$) and is correlated with the daily amount spent and views generated (Pearson $R \in [0.87, 0.88]$ $P < .001$), as shown in Figure \ref{fig:ts-general}. A similar result holds for the number of unique ads active each day (for which we omit the figure) -- with a peak of over 1 k ads created and over 5 k active ads a few days before the elections.

\begin{figure}[!t]
    \centering
   \includegraphics[width=\linewidth]{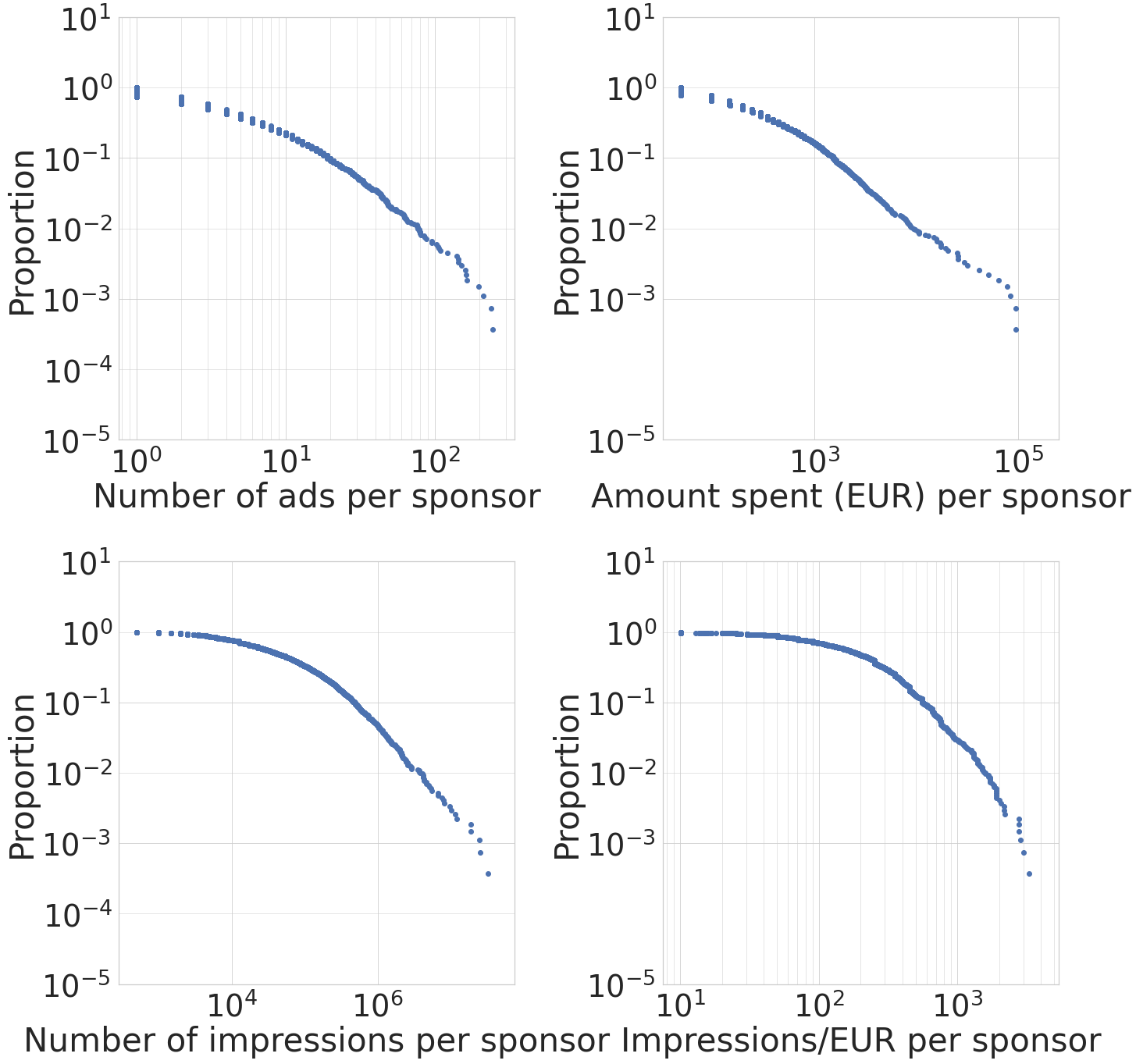}
    \caption{Distributions of the number of ads, the amount spent, number of impressions and impressions/EUR for each sponsor in the dataset. Both axes are in logarithmic scale.}
    \label{fig:sponsor-distributions}
\end{figure}

A similar analysis for sponsors shows that, on average, they created 8.5 ads during the period of analysis (median: 3 ads), with the most prolific sponsor creating 261 ads. Each sponsor spent on average 9400 EUR (median: 199 EUR), with a maximum value of over 166 k EUR spent by a single sponsor. Sponsors generated 27 k impressions on average (median: 4 k impressions), with the most viewed sponsor generating over 42 M impressions. Sponsors generated 270 impressions for each EUR spent on average (median: 186 impressions). Similarly to individual ads, these metrics follow an exponential-like distribution as shown in Figure \ref{fig:sponsor-distributions}. We find a very strong correlation between money spent and impressions generated at the sponsor level (Pearson $R=0.94$, $P < .001$), and a weak correlation between these metrics and the number of ads generated (Pearson $R \in [0.46, 0.49]$, $P < .001$). The number of unique sponsors active each day follows a pattern similar to those mentioned above, and we omit the figure. A manual analysis of most active sponsors reveals that they are associated with main political parties and candidates, which we analyze in a grouped fashion in the next sections. 

\textbf{Remarks:} We measured the extent to which Facebook and Instagram were used for online advertising in the run-up to the 2022 Italian general election, finding an increasing trend toward election day (September 25th). On average, ads sponsored in this period were paid 111 EUR, generating 32 k impressions over a period of 4.5 days, with a cost per thousand impressions of 4 EUR. The average sponsor active during the period of analysis created around 8.5 ads, spending 9.4 k EUR and generating 27 k impressions. This first analysis shows that a single ad can be delivered on these platforms at a very low-cost and (potentially) reach thousands of individuals.

\subsection{Spending activity and engagement of political sponsors}

\begin{figure*}[!t]
    \centering
    \includegraphics[width=\linewidth]{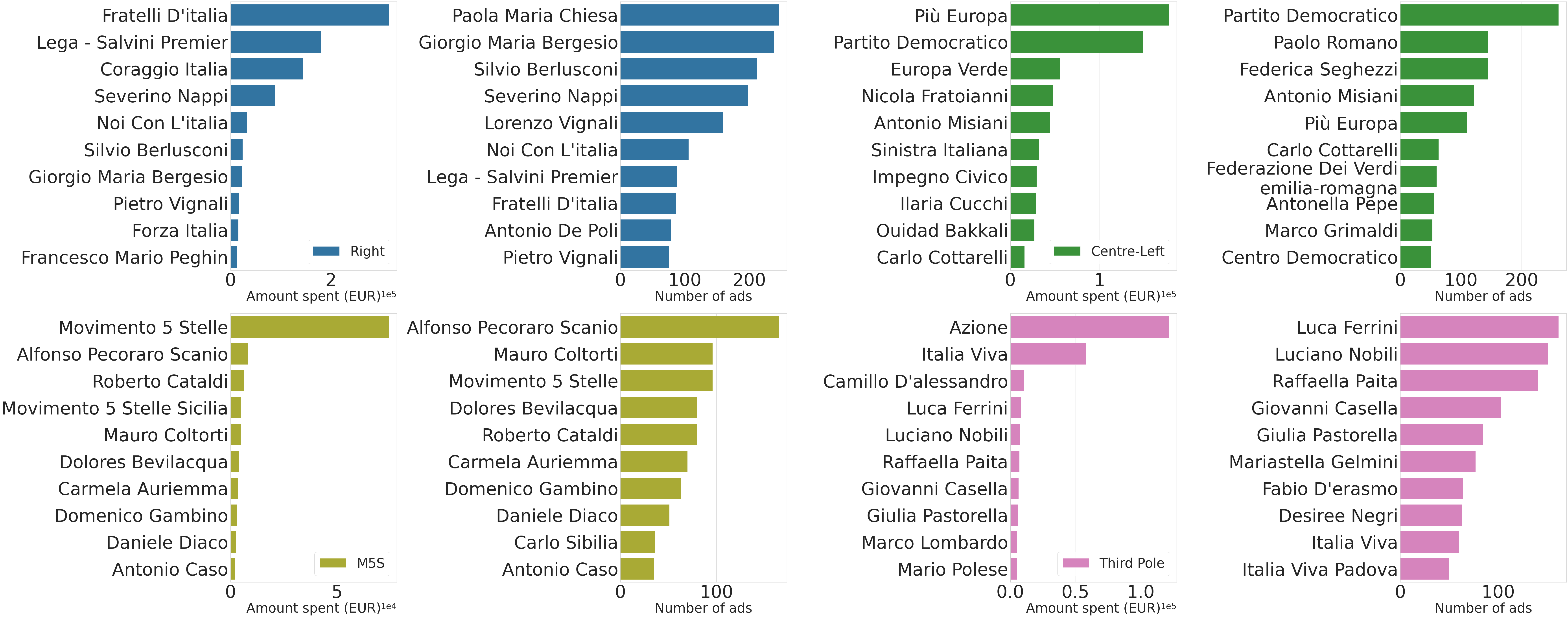}
    \caption{Top 10 sponsors in terms of amount spent and number of created ads, for each political group.}
    \label{fig:barplot-parties}
\end{figure*}

In Figure \ref{fig:barplot-parties}, we show the top 10 sponsors for each political coalition in terms of amount of money spent (we omit impressions because they are strongly correlated) and number of ads created. We can notice that in all cases parties spent a much higher amount of money than individual candidates, with this discrepancy being particularly high in the case of M5S and the Third Pole. Interestingly, parties that obtained a small percentage of votes in the Centre-Left coalition (namely "Più Europa" and "Europa Verde") spent an amount comparable to the main party ("Partito Democratico"). Also, Berlusconi's party ("Forza Italia") spent a considerable smaller amount of money compared to "Lega", yet their share of votes was similar.

\begin{figure}[!t]
    \centering
    \includegraphics[width=\linewidth]{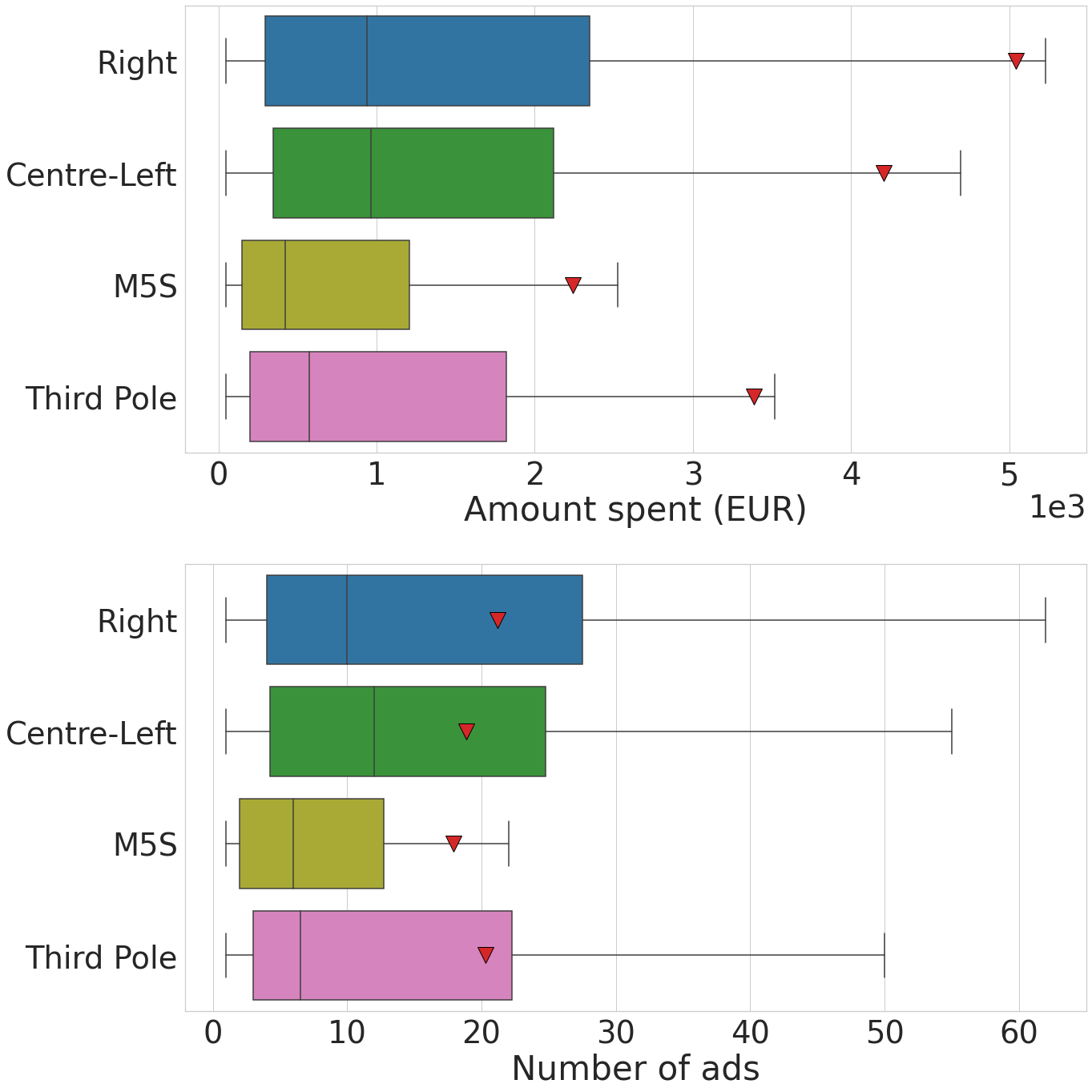}
    \caption{Distributions of the amount spent \textbf{(top)} and number of ads created (bottom) at the sponsor level, for each political coalition. Median values for the \textbf{top} panel are: Right = 940.5, Centre-Left = 966.25, M5S = 420, Third Pole = 527.5. Median values for the \textbf{bottom} panel are: Right = 10, Centre-Left = 12, M5S = 6, Third Pole = 6.5. Triangles indicate the mean value of distributions, and we do not visualize outliers.}
    \label{fig:boxplot-sponsors}
\end{figure}

In Figure \ref{fig:boxplot-sponsors} we show the distributions of amount spent and number of ads created at the sponsor level, for each political coalition. Running a two-sided Mann-Whitney test on the distributions of the two metrics, we compare Right versus Centre-Left affiliated sponsors, and similarly M5S versus Third Pole sponsors. We find that, in the former comparison, sponsors spent a comparable amount of money ($P > .1$ ) and created a similar number of ads ($P > .1$ ); in the latter comparison, the Third Pole spent a larger amount of money ($P < .001$ ) but created a similar amount of ads ($P > .1$ ). We report median values in the caption of the figure.

\textbf{Remarks:} We analyzed the advertising campaign of political coalitions in terms of ads created, money spent and impressions generated aggregating on the entire period of analysis. We first highlighted the most prolific sponsors, finding that official parties spent a significantly larger amount of money than individual candidates. Analyzing political coalitions as a whole, we found that the two most performing groups were the Right and Centre-Left coalitions, which respectively ranked 1$^{st}$ and 2$^{nd}$ at the elections. Among the other two competitors, M5S and the Third Pole, the latter was significantly more active than the former, despite the smaller share of votes obtained in comparison. This analysis confirms the potential of Meta Ad library as a monitoring tool for political advertising, thus allowing for transparency and accountability of actors promoting specific political agendas with marketing campaigns. 

\subsection{Temporal patterns in the advertising activity of political coalitions}

\begin{figure}[!t]
    \centering
    \includegraphics[width=\linewidth]{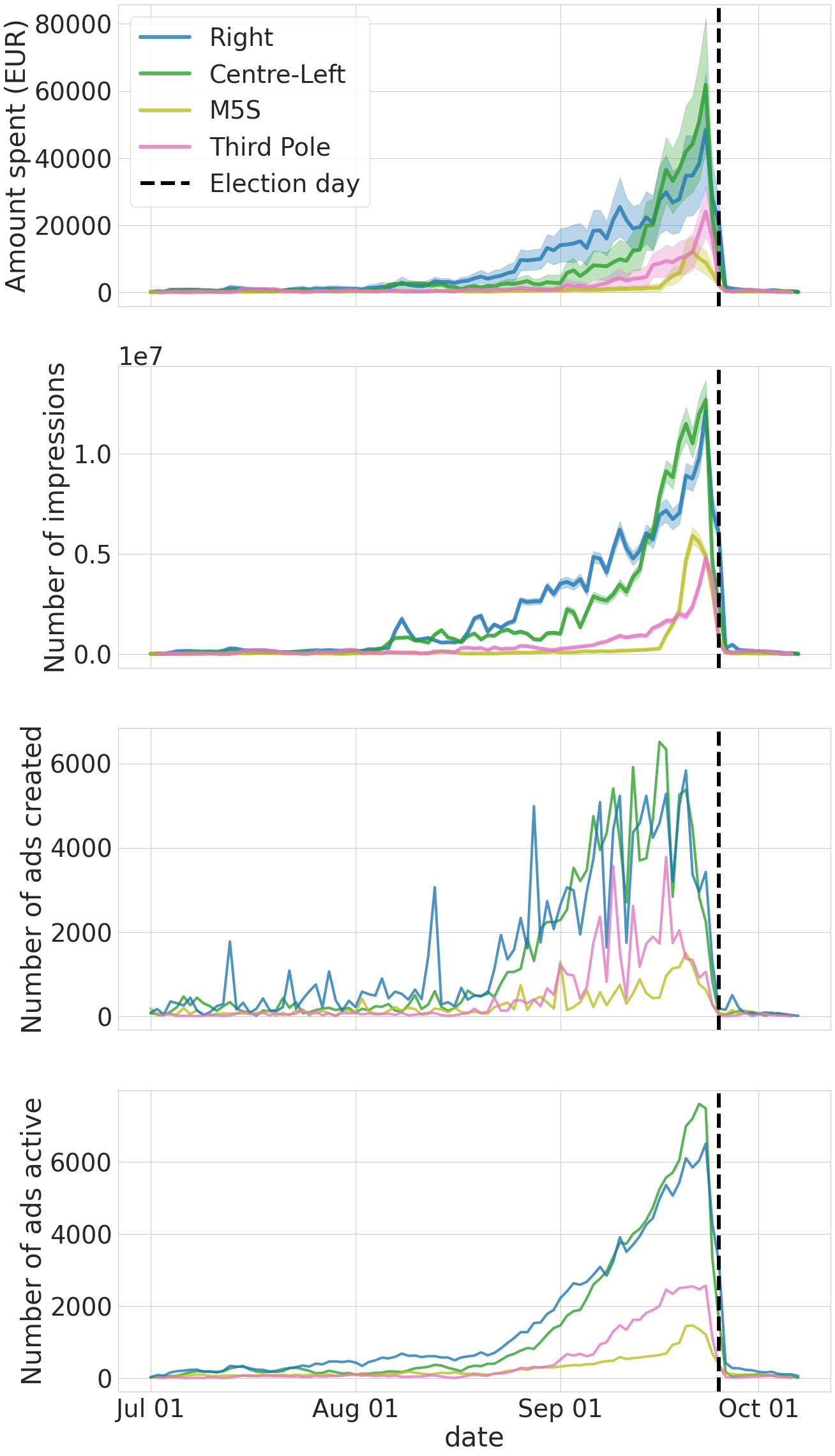}
    \caption{Time series of the amount spent, impressions generated, ads created and active on each day, for each political coalition. In the two upper panels the dashed area represents lower and upper bounds as provided by Meta Ad Library API, with the solid line corresponding to the mean value, aggregated at the group level.}
    \label{fig:ts-parties-amount-engagement}
\end{figure}

In this section, we analyze the temporal activity of political groups in terms of amount spent, impressions generated, ads created and active. We compute aggregated metrics for each day by grouping together ads pertaining to sponsors that we matched to any of the four main political groups, as described in the methods section.

As shown in Figure \ref{fig:ts-parties-amount-engagement}, the political advertising activity of all political groups exhibits an increasing trend towards election day (Mann-Kendall $ P> .001$  in all cases), similar to the general trend highlighted in the previous section, and there are strong correlations among groups (Pearson $R \in [0.74, 0.95]$, $P < .001$). The main coalitions, namely Right and Centre-Left, were the most active and generated most of the impressions; in particular, we notice that the Right coalition started to spend a larger amount of money compared to the Centre-Left coalition towards the end of August, whereas we observe that the latter was more active in the last few weeks of the election period, overtaking the former in terms of amount spent and views generated. In what regards the other two coalitions, Third Pole was apparently more active than M5S, especially in the final period, but generated fewer impressions. Both groups spent a significantly smaller amount of money compared to the Right and Centre-Left coalitions, which were also those that obtained most of the votes. Finally, the number of ads created and active on each day exhibit trends similar to panel Figure \ref{fig:ts-general}.

\begin{figure}[!ht]
    \centering
    \includegraphics[width=\linewidth]{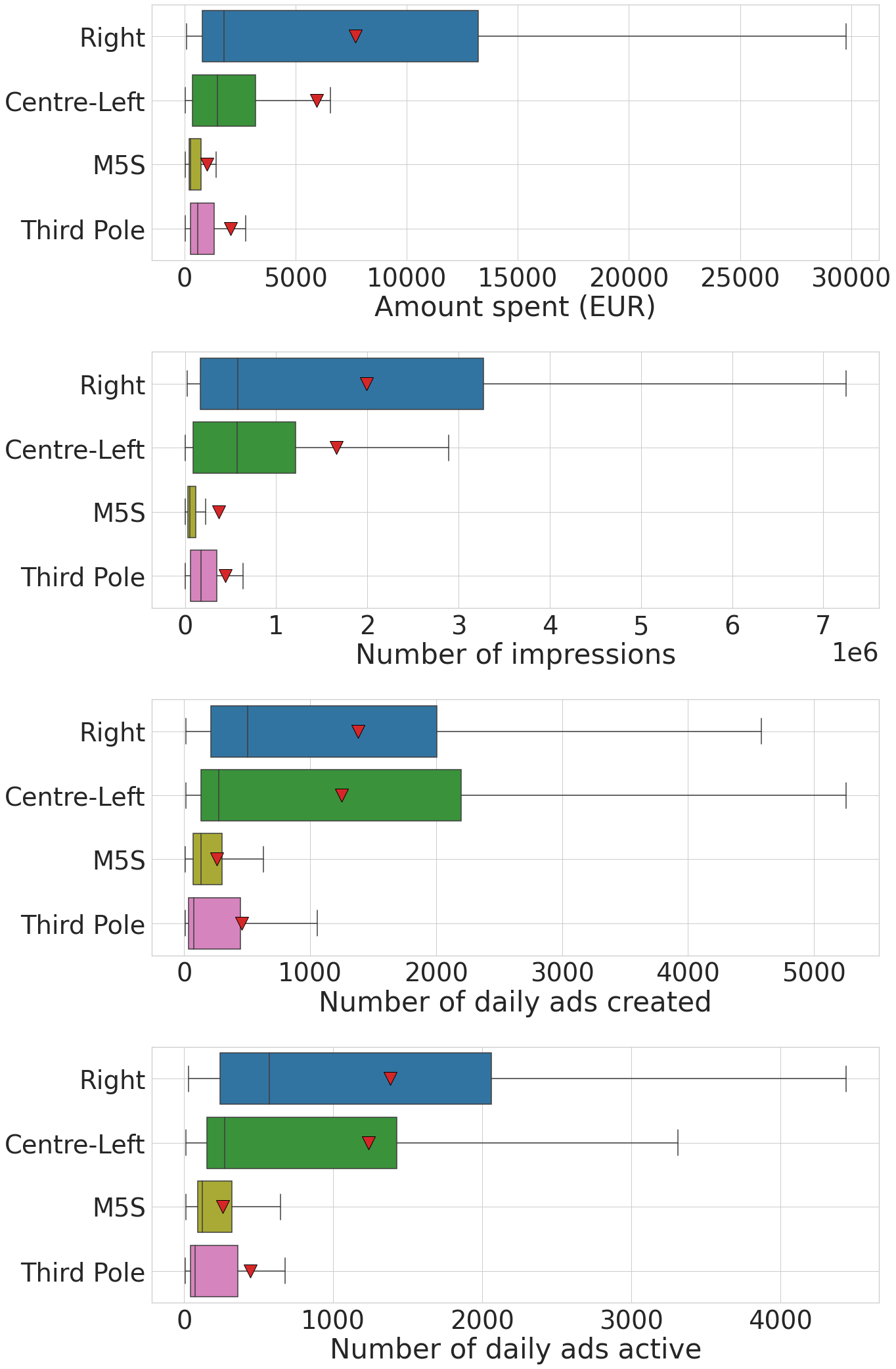}
    \caption{Distributions of the amount spent, impressions generated, ads created and active on each day, for each political coalition. Triangles indicate the mean value of distributions, and we do not visualize outliers.}
    \label{fig:boxplot-parties}
\end{figure}

These results are confirmed in Figure \ref{fig:boxplot-parties}, where we show the distributions of daily values for the following metrics: amount spent, number of impressions generated, number of ads created and active. Given the comparable values observed in the Figure, we used two-sided Mann-Whitney tests to assess differences between Right and Left coalitions, finding that on the average day:
\begin{itemize}
    \item the Right coalition spent more than the Centre-Left (Right median: 2,124, Centre-Left median: 1,619, $P < .01$)
    \item the Centre-Left coalition generated more views than the Right (Right median: 625,647, Centre-Left median 670,126, $P < .01$)
    \item the Right coalition created more ads than the Centre-Left coalition (Right median: 537, Centre-Left median: 303, $P < 0.05$)
    \item the Right coalition had more active ads than the Centre-Left coalition (Right median: 576, Centre-Left median: 285, $P < 0.05$)
\end{itemize}
and similarly between M5S and Third Pole, finding that on the average day:
\begin{itemize}
    \item the Third Pole spent more than M5S (Third Pole median: 659, M5S median: 267, $P < .001$)
    \item the Third Pole generated more views than M5S (Third Pole median: 185,496, M5S median: 53,726, $P < .001$)
    \item the M5S created more ads than the Third Pole (Third Pole median: 78, M5S median: 135, $P < .001$)
    \item the M5S had more active ads than the Third Pole (Third Pole median: 72, M5S median: 120, $P < .005$)
\end{itemize}

\textbf{Remarks:} We investigated the presence of temporal patterns in the advertising campaigns of different coalitions, finding inter-group correlations and trends that confirmed the higher prevalence of ads sponsored by the Right and Centre-Left coalitions. In particular, despite a larger amount of money spent and ads created, on the average day the Right coalition generated slightly fewer views than the Centre-Left coalition, most likely due to the increased advertising activity of the latter group in the weeks preceding the election. The Third Pole spent more and generated more views than the M5S, which, however, created more ads on an average day. These findings suggest that political advertising campaigns can be monitored over time in order to detect interesting patterns and understand strategies put in place by different political entities.

\subsection{Geographical and demographic patterns in the advertising activity of political coalitions}
In this section, we analyze the demographic and geographical characteristics of the political advertising campaigns of different coalitions. We consider all ads by political groups that contain a breakdown of the amount spent and the number of impressions generated specifying at least one Italian region, and different age cohorts.\footnote{Considering ads that are uniformly distributed over the country and age cohorts does not affect the results.}

\begin{figure}[!t]
    \centering
\includegraphics[width=\linewidth]{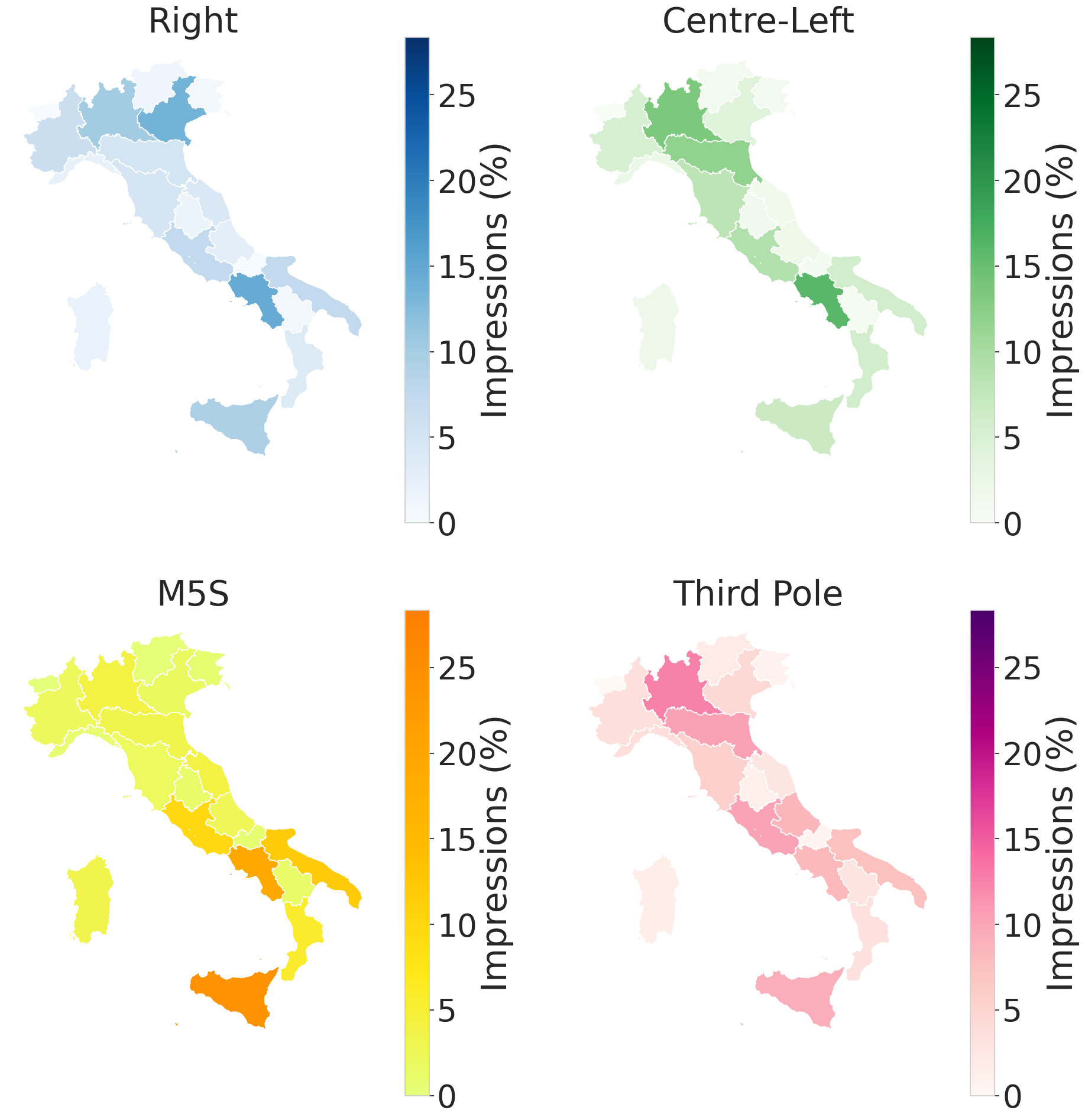}
    \caption{Distribution of the proportion of impressions generated by each political coalition in all Italian regions, normalized by the total sum spent overall by each group. The max value in the colorbar corresponds to the maximum proportion across all coalitions.}
    \label{fig:heatmap-spending}
\end{figure}

In Figure \ref{fig:heatmap-spending} we show the proportion of the impressions generated in each region by different coalitions, normalizing by the total number generated by each group. We omit the same figure for the number of ads created as the metrics are highly correlated (Pearson $R = 0.79$, $P < .001$ for all groups). We do not consider the amount spent for the following reason: the Ad Library provides information about the location and demographics of the individuals that ultimately saw ads but not of the audience that was targeted, thus the statistics cannot be really considered for determining campaign targeting \cite{brodnax2022home}. We see a striking similar geographical distribution in the audience of Centre-Left and Third Pole, in line with their overlapping political agenda\footnote{Matteo Renzi, the current secretary of Italia Viva, is the former secretary of Partito Democratico.}; in fact, the amount of impressions generated in each region by both groups is highly correlated (Pearson $R = 0.81$, $P < .001$). Inter-group correlations are also significant (Pearson $R \in [0.56, 0.77]$, $P < .001$). We also observe that M5S is the group that received most of the impressions in southern regions, which is also where the party obtained most of its elected representatives. 

\begin{figure}[!t]
    \centering
    \includegraphics[width=\linewidth]{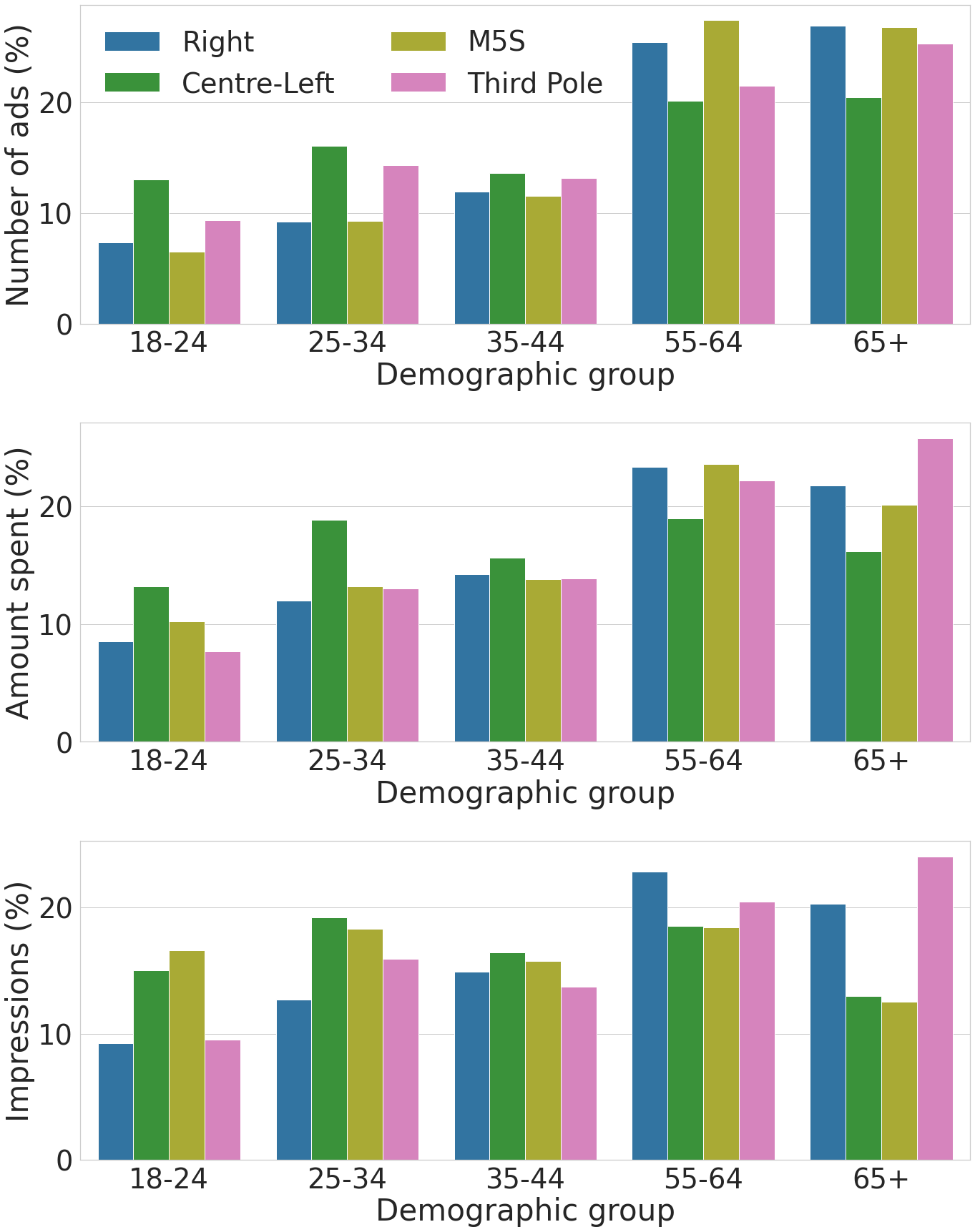}
    \caption{Demographic distribution for the number of ads, amount spent and number of impressions generated, for each political coalition. }
    \label{fig:demographics}
\end{figure}

In Figure \ref{fig:demographics} we show the demographic breakdown of political advertising of different groups in terms of the proportion of ads created, amount of money spent, and impressions generated. We observe that in general political coalitions targeted different age cohorts in a similar fashion, with much attention devoted to the older population, most likely because the younger generations tend to participate less in the democratic process. This is particularly noticeable for the amount spent and impressions generated by the Right and Third Pole groups in the youngest and oldest age cohorts (18-25 vs 65+), whereas Centre-Left and M5S targeted the Italian population more uniformly.

\begin{figure}[!t]
    \centering
    \includegraphics[width=\linewidth]{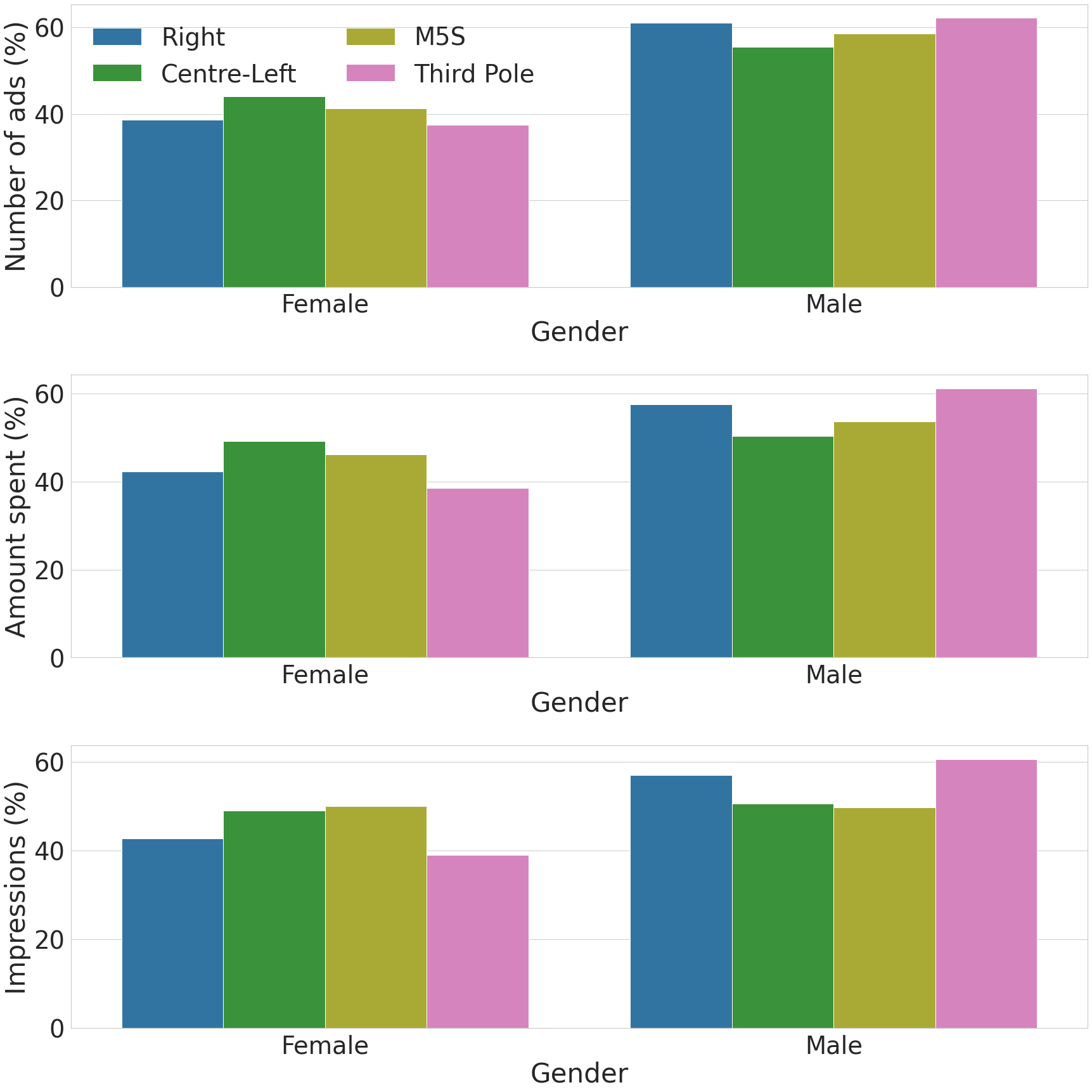}
    \caption{Gender distribution for the number of ads, amount spent and number of impressions generated, for each political coalition. }
    \label{fig:gender}
\end{figure}

Finally, as shown in Figure \ref{fig:gender}, we report slightly larger attention towards men than women in the advertising activity of all coalitions for the Right and Third Pole coalitions, whereas Centre-Left and M5S apparently targeted both genders.

\textbf{Remarks:} We investigated geographical and demographic patterns in the political campaigns of different coalitions. We found inter-group correlations that reflect the extent to which different groups reached a similar audience in different geographical areas of the peninsula, with Centre-Left and the Third Pole being particularly similar, most likely due to an overlapping agenda. The M5S reached most of its audience in the Southern regions, which paid back at the elections. We also showed that all coalitions focused more on the eldest cohorts and that some groups targeted more men than women. This analysis indicates effective ways to ascertain the political agenda of candidates and parties in terms of the demographics of the targeted audiences.

\section{Conclusions}
\subsection{Contributions}
We studied the usage of Meta platforms, namely Facebook and Instagram, to advertise political content in the run-up to the 2022 Italian general election. Leveraging Meta Ad Library, we collected more than 20 k unique ads advertised over a period of three months, with a total amount of 4 M EUR spent on political advertising. We estimated that most of the ads had a cost per thousand impressions of 4 EUR while most sponsors created less than 10 ads, and that the amount spent on advertising was strongly correlated with the number of views generated. Employing a manual procedure to label political ads, we highlighted similarities and differences in the advertising campaign of the main political coalitions, from different perspectives. We found that the two most voted coalitions, namely the Right and the Centre-Left coalitions, were also those that spent the most in online advertising, and generated most impressions. The former, however, obtained a larger proportion of seats compared to the latter, despite the similar amount spent on advertising. The two other main competitors, M5S and the Third Pole, spent a significantly smaller amount of money, thus generating fewer impressions. We observed that coalitions exhibited an increasing trend of political advertising toward election day (September 25th), with the Centre-Left coalition being particularly active in the last few weeks. Finally, we highlighted geographical and demographic patterns in the campaigning strategy of different coalitions.

\subsection{Limitations}
Our work does not come without limitations. First, as highlighted in \cite{le2022audit,silva2020facebook}, Meta might not accurately label all political ads as such, and our collection might be missing some data. Besides, some ads might not be related to the elections despite matching related keywords, although we address this issue by manually labeling sponsors. Our data measures impressions generated by ads that do not necessarily indicate the number of unique individuals reached by the sponsored content, as the same ad might have been shown to the same person more than once, and we cannot ascertain the extent to which audiences overlap in different campaigns. Also, the library does not provide information about ads that are placed to custom audiences and/or targeted explicitly by keywords \cite{brodnax2022home}. Moreover, Instagram and Facebook user-based might not be fully representative of the Italian population \cite{pew2019social}, and there might also be mismatches between the actual demographics of users and those inferred by Meta advertising algorithms \cite{grow2022facebook}. Finally, we note that we did not perform an in-depth analysis of the content sponsored by different political groups in order to study their political communication. A basic topic analysis based on the most frequent keywords and weighted log odds did not reveal particular differences in the content sponsored by groups, as most of the ads invited to vote for specific candidates and parties. We, therefore, leave a more detailed analysis in this direction for future work.

\subsection{Discussion and Future Work}
Our findings confirm the relevance of online advertising put in place on social platforms during political elections, which is particularly relevant given that the micro-targeting feature has caused a wide backlash in the past \cite{capozzi2021clandestino}, as consumers might not be always able to distinguish paid from unpaid content \cite{johnson2018analyzing}. It also adds to a stream of literature that leverages publicly available data from social media to investigate the potential impact of online social media on real-world phenomena. Until recently it was challenging to monitor political advertisement campaigns put in place on social media platforms, and we believe that Meta Ad Library offers researchers endless possibilities to explore the data. There is still scarce evidence on whether digital advertising campaigns might actually affect voters' choice \cite{coppock2022does, aggarwal20232}, e.g., we found that the Right and Centre-Left coalitions spent a similar amount but the latter obtained a larger number of seats.
Future research might consider a number of different directions. We did not disentangle similarities and differences between Facebook and Instagram, and researchers could further investigate how political communication strategies differ on the two platforms, given the demographic differences in their user basis. They could also focus on platforms such as TikTok, which are gaining consensus among young people and might attract political campaigns in the future. Another possibility is to focus more on content-based analyses of the ads sponsored by different parties, in order to better understand the topics and themes on which politicians built their political campaigns. This includes potentially misleading content and narratives delivered by political figures to persuade voters. Researchers could also investigate the online advertising campaigns of fringe and extreme candidates, most of which were promoting anti-establishment and conspiratorial narratives. Finally, future research could leverage Meta Ad library in a similar fashion to study other contexts where digital advertising might play an important role in shaping individuals' opinions and actions.

\subsection{Ethical Concerns}
We analyze and process data in accordance with Meta terms of service and use. Ads were collected through a public API and no individual users were de-identified nor harmed in the process. We provide access to the IDs of ads in the spirit of transparency and reproducibility, thus allowing others to retrieve the dataset in accordance with Meta data-sharing policies, which require a few identification steps in order to access its Ad Library through the API (whereas the interactive search console is available to anyone).

\section{Acknowledgments}
Work supported in part by PRIN grant HOPE (FP6, Italian Ministry of Education).

\bibliographystyle{ACM-Reference-Format}
\bibliography{bib}

\end{document}